\documentclass{emulateapj}
\usepackage{apjfonts}

\newcommand{\tev}{TeV~\-J2032+4130}
\newcommand{\ls}     {LS~5039}

\newcommand{\prp}    {${\rlap.}^{\prime}$}
\newcommand{\grp}    {${\rlap.}^{\circ}$}
\newcommand{\pri}    {${\rlap.}^{\prime \prime}$}
\newcommand{\rl}     {${\rlap.}^{s}$}

\newcommand{\lsi}    {LSI+61$^{\circ}$303}

\newcommand{\ltsima} {$\; \buildrel < \over \sim \;$}
\newcommand{\simlt}  {\lower.5ex\hbox{\ltsima}}            
\newcommand{\gtsima} {$\; \buildrel > \over \sim \;$}
\newcommand{\simgt}  {\lower.5ex\hbox{\gtsima}}            

\slugcomment{Accepted by ApJ {\it Letters}}

\shorttitle{Radio sources in the field of TeV~J2032+4130}
\shortauthors{Paredes et al.}

\begin{document}


\title{The population of radio sources in the field
of the unidentified $\gamma$-ray source \tev}


\author{
Josep M. Paredes,\altaffilmark{1}
Josep Mart\'{\i},\altaffilmark{2}
Ishwara Chandra C. H.,\altaffilmark{3}
Valent\'{\i} Bosch-Ramon\altaffilmark{4}
}
\altaffiltext{1}{Departament d'Astronomia i Meteorologia, Facultat de F\'{\i}sica, Universitat
de Barcelona, Mart\'{\i} i Franqu\`es, 1, 08028 Barcelona (Spain), jmparedes@ub.edu}
\altaffiltext{2}{Departamento de F\'{\i}sica, Escuela Polit\'ecnica Superior, Universidad de Ja\'en,
Las Lagunillas s/n, 23071 Ja\'en (Spain), jmarti@ujaen.es}
\altaffiltext{3}{National Center for Radio Astrophysics, TIFR, P. B. No. 3, Ganeshkhind, Pune - 7, India, ishwar@ncra.tifr.res.in}
\altaffiltext{4}{Max Planck Institut f\"ur Kernphysik, Heidelberg 69117 (Germany), vbosch@mpi-hd.mpg.de}









\begin{abstract}

\tev\ is the first extended very high energy gamma-ray source, which has remained enigmatic since its discovery, due to the lack of
identification. We report here deep radio observations covering the \tev\ field and revealing for the first time an extended and diffuse
radio emission, as well as a remarkable population of compact radio sources. Some of these radio sources are in positional coincidence with X-ray and
optical/IR sources. Future follow up studies of these new radio sources will likely contribute to solve the mystery of this extended
unidentified TeV source. 
\end{abstract}





\section{Introduction}

The new generation of Cherenkov telescopes has revealed the existence of a population of very high energy gamma-ray sources in the Milky Way
\citep{aha2005a,aha2006}. Some of these sources are well identified and are compact sources like \ls ~\citep{aha2005b} or \lsi
~\citep{alb2006},   whereas others still remain without known counterpart. Among the latter there are some extended TeV
sources \citep{aha2002,aha2005c} for which all attempts to find their counterparts at other wavelengths have failed, and are being considered
a new population of galactic sources. Because the lack of detection of a low energy counterpart it has been proposed that  hadronic
processes instead of leptonic ones could be behind the TeV emission  (e.g.\cite{t2004}).

The so far unidentified source \tev, discovered with the stereoscopic 
High Energy Gamma Ray Astronomy (HEGRA) array of imaging Cherenkov telescopes 
in the direction of the Cygnus OB2 star association \citep{aha2002}, 
is perhaps the most representative within the group of TeV sources with unknown counterpart. 
Other examples are HESS~J1303$-$631 ~\citep{aha2005c}, the second unidentified TeV source, 
and the unidentified sources found in the HESS galactic plane survey ~\citep{aha2005a}. 
The identification of any of these sources, by locating its counterpart at radio, optical or infrared wavelengths, would improve our knowledge 
of the emission processes at very high energies and would help to solve the problem of the unidentified TeV sources.
TeV J2032+4130 is extended, with a radius of 6\prp 2, being the center of gravity (CoG) of TeV photons 
located at $\alpha_{\rm J2000.0}$ = $20^h 31^m$57\rl 0 and $\delta_{\rm J2000.0}$ = $+41^{\circ} 29^{\prime}$56\pri 8 with arc-minute accuracy. Its emission exhibits a hard spectrum with a steady flux of about a few 
$10^{-12}$ erg cm$^{-2}$ s$^{-1}$ ~\citep{aha2005d}. The large angular size of the source, 
its steady flux and the location in the galactic plane point strongly to a galactic nature. 
The absence of any evident counterpart led \citet{aha2002} to propose as possible origins of the 
TeV emission the star association Cygnus OB2 or a jet-driven termination shock.
Hadronic interactions within the innermost region of the winds of O and B stars can also produce significant gamma-ray luminosities 
at TeV energies, compatible with those found in TeV J2032+4130 \citep{t2004,ds2006}.
 
Despite the intensive observational campaigns undertaken for this source at different wavelengths,  no obvious counterpart at radio, optical
nor X-ray energies has been reported up to now, leaving \tev\ presently unidentified (\cite{butt2003,butt2006,mu2003}). Here we report the
discovery of a population of radio sources well within the $1\sigma$ radius of the \tev\ extended emission, 
presenting some of these radio sources 
optical/IR and X-ray counterparts. One or several of these objects, including both a extended diffuse component and compact radio sources,
could be related to the  very high energy emission. Their observational properties and nature are presented and discussed in the following
sections.

\section{Observations and analysis}

\subsection{Radio}

Most of the data presented in this work were obtained through observations performed with the {\it Giant Metrewave Radio Telescope} (GMRT)
of the National Centre for Radio Astrophysics (NCRA) in Khodad (India), during 2005 July 9, 2005 September 1 and 2006 July 18. We observed
at the 610 MHz frequency (49 cm wavelength) in spectral line mode with 128 channels covering a 16 MHz bandwidth. The calibration of
amplitude and bandpass was achieved by observing 3C~286 and 3C~48, while phase calibration was perfomed through repeated scans on the nearby
phase calibrator J2052+365. Correction of the GMRT flux densities for the increase in the sky  temperature in the \tev\ direction was also
taken into account. In addition, we also used archive data obtained during 2003 April 29 using the {\it Very Large Array} (VLA) of the
National Radio Astronomy Observatory (NRAO) in New Mexico (USA). The VLA was in its most compact D configuration and the observations were
carried out at the 1.465 and 4.885 GHz (20 and 6 cm wavelength, respectively). The VLA amplitude calibrator was 3C~48,  and J2052+365 and also
J2007+404 were used as phase calibrators. Both the GMRT and VLA data were processed using standard procedures within the AIPS software
package of NRAO.

The results of the radio observations are presented in Figs. \ref{fig1} and \ref{fig2}. The left panel of Fig. \ref{fig1} contains a wide
field map of the \tev\ region obtained by mosaicing four individual pointings with the VLA at 20~cm. Extended and diffuse radio emission is
clearly seen inside the \tev\ $1\sigma$ radius and closely surrounding the CoG of TeV photons, indicated by the central cross. The overall
morphology of the main extended emission features in the field is also detected in GMRT maps (not shown here) retaining short spacings, confirming their shape and extension. A spectral index image (not shown here) created using matching beam VLA maps at 20 and 6 cm strongly suggests
that there is a diffuse emission half surrounding the CoG of the \tev\ source (red dashed arc in Fig. \ref{fig1} left) 
with non-thermal spectral index $\alpha \simeq -1$, defined as $S_{\nu} \propto \nu^{\alpha}$. 
We note that the diffuse emission seen 
well outside the TeV source circle is thermal, hinting to a
differentiated origin for the non-thermal diffuse radiation within the TeV source circle and to a physical link between the diffuse radio
mission and the TeV source.
Another remarkable object in the field is a prominent double-lobe radio galaxy 
whose core is at $\alpha_{\rm J2000.0}$ = $20^h 32^m 01$\rl 7 and $\delta_{\rm J2000.0}$ = $+41^{\circ} 37^{\prime} 22^{\prime\prime}$, 
with an integrated flux density of $91\pm1$ mJy at the 20 cm wavelength. 

The right panel of Fig. \ref{fig1} illustrates a zoomed VLA map at 6 cm of a region close to the TeV CoG.  At this shorter wavelength we can
clearly detect here two radio sources with different morphology.  The northern one (VLA-N) is practically compact, 
while the sourthern VLA source (VLA-S) is not consistent with being point-like.

The most compact source sensitive radio view of the \tev\ field comes nevertheless from the final 610 MHz map
combining all different GMRT observing epochs.
This map is presented in grey scale plot in the central panel of Fig. \ref{fig2} and provides the
deepest radio image with arc-second detail ever obtained so far of the \tev\ position.
Extended emission has been removed by not including visibilities corresponding
to baselines shorter than 1 k$\lambda = 490$~m. The field shown covers the entire $1\sigma$ radius
and at least six compact radio sources are detected with peak flux density
above five times the rms noise of about 90 $\mu$Jy beam$^{-1}$.
Their observed properties are listed in Table \ref{sources}. 

Finally, we will also mention that a short VLA exploratory time proposal in A configuration was also 
conducted by us at the 3.5~cm wavelength on 2006 April 1 and 3rd. As a result, a compact (angular size $\leq$0\pri 3)
radio source  was detected with a flux density of $0.24\pm0.03$ mJy located at 
$\alpha_{\rm J2000.0}$ = $20^h 31^m 52$\rl 686 and $\delta_{\rm J2000.0}$ = $+41^{\circ} 30^{\prime}$54\pri 56. 
This corresponds to the small cross in the right panel
of Fig. \ref{fig1} (within the radio contours of the mentioned VLA-S source), with a {\it Core} label for reasons discussed below.

\subsection{Infrared}

The field of \tev\ was observed in the near infrared $Ks$-band (2.2 $\mu$m) using the
Calar Alto 3.5~m telescope and the OMEGA2000 camera on 2005 April 29th. Another $Ks$-band image
of the field was taken on 2006 July 7th using the 4.2~m William Herschel Telescope (WHT) and the LIRIS imager/spectrograph.
These observations were processed in a standard way as described in \citet{m2006} with the IRAF package of the
National Optical Astronomy Observatories (NOAO), including sky background subtraction and flat fielding.
The left and right panels of Fig. \ref{fig2} display part of these observations showing the identification
of near infrared counterparts to some of the radio sources reported in this paper. The identification is on the basis
of astrometric coincidence within a few tenths of an arc-second.

\section{Results and discussion}

An important consideration concerning the new radio sources reported here is their coincidence with X-ray
emitters in the field previously reported by other authors \citep{butt2003,butt2006,mu2003}. Hereafter, X-ray sources
will be referred to using the  {\it Chandra} \citet{butt2006} identification numbers. 
We will discuss this issue for the three most relevant objects reported in this work, namely,
the VLA-N source (or GMRT source \#3), the VLA-S source (undetected by the GMRT)
and the GMRT source \#5. VLA-N and VLA-S are well detected separately only at 6 cm, but an estimate of their individual 20 cm flux densities
is feasible by fitting two Gaussian components with fixed positions.

\begin{figure*}
\includegraphics[angle=0,scale=0.90]{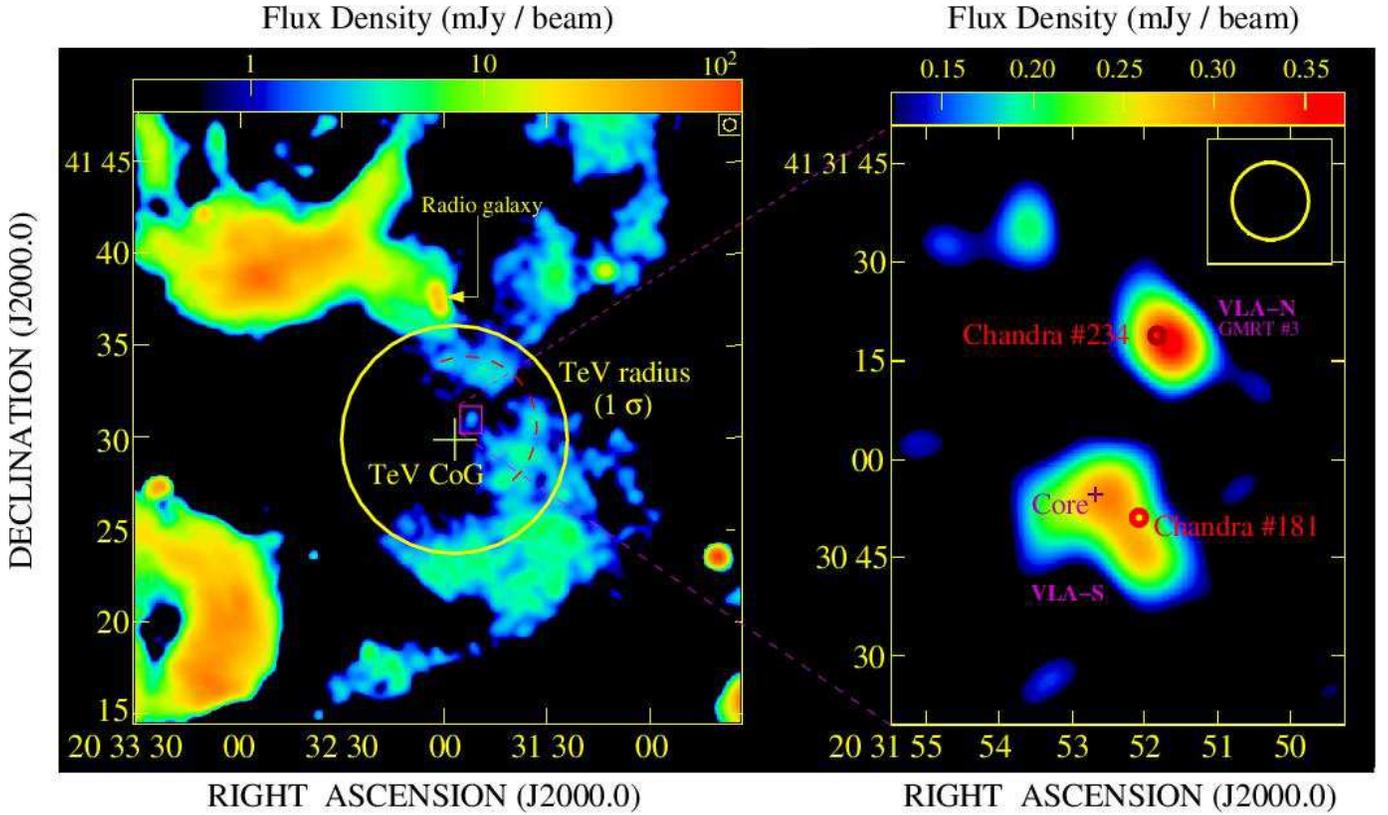}
\caption{{\bf Left.} Large scale VLA mosaic, at the 20 cm wavelength (1.4 GHz) in D configuration and
uniform weight, of the unidentified $\gamma$-ray source \tev. 
The CoG of TeV emission and its statistical error are
indicated by the central cross, while the big yellow circle ilustrates
the $1\sigma$ radius of the TeV extended emission. Inside it, diffuse radio emission is
clearly seen. The red half arc traces the extended emission around the TeV CoG position.
Color scale goes logarithmically
from above 0 to 126 mJy. The synthesized beam is shown at the top right corner
as an 40\pri 7$\times$40\pri 2 ellipse, with position angle of 15\grp 3. 
{\bf Right.} Zoomed map of an interesting area inside the TeV extended emission circle
where a complex of VLA radio sources has been detected at the 6 cm wavelength
(4.8 GHz).
Here, the red circles indicate the location of {\it Chandra} X-ray sources in the field, labeled 
according to the X-ray identification number given by  \cite{butt2006}. The {\it Chandra} source \#234, at $\alpha_{\rm J2000.0}$ = $20^h 31^m$51\rl 84 and $\delta_{\rm J2000.0}$ = $+41^{\circ} 31^{\prime}$18\pri 84, is a hard X-ray source and is one of the strongest sources in the field of TeV~2032+4130. This source has also been detected at 610 MHz (source \#3 in Table~1). The {\it Chandra} source \#181, $\alpha_{\rm J2000.0}$ = $20^h 31^m$52\rl 08, $\delta_{\rm J2000.0}$ = $+41^{\circ} 30^{\prime}$51\pri 12, is a weak source.
The small cross shows the location, $\alpha_{\rm J2000.0}$ = $20^h 31^m 52$\rl 686 and $\delta_{\rm J2000.0}$ = $+41^{\circ} 30^{\prime}$54\pri 56, of a compact radio core detected
at 3.5 cm during a follow up run with the  VLA in A configuration (not shown here). 
Uniform weight was used to create
this 6 cm map with a synthesized beam of 11\pri 9$\times$11\pri 8 with position angle of $-$13\grp 0. 
The color scale goes linearly from nearly zero to 0.37 mJy beam$^{-1}$.
\label{fig1}}
\end{figure*}

\begin{figure*}
\includegraphics[angle=0,scale=0.900]{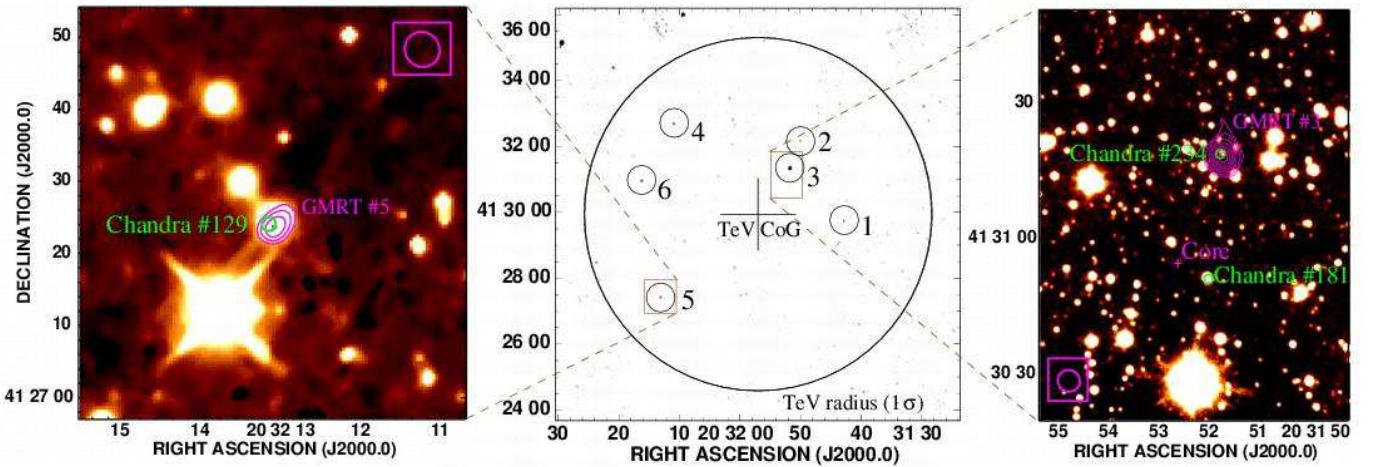}
\caption{{\bf Center.} The central panel shows our deep GMRT map of the \tev\ field at the
49 cm wavelength (610 MHz). The central cross and the big circle have the same meaning
as in Fig. \ref{fig1}. Up to 6 compact sources are detected above a $5\sigma$ peak flux density level
inside the TeV extended emission circle.
Those with both X-ray and near infrared counterparts are shown in detail in the panels aside.
Uniform weighting, excluding baselines shorter of 1k$\lambda$, has been used to produce this GMRT map with a synthesized beam
of 5\pri 0$\times$4\pri 8 and position angle of 33\grp 0. The grey scale is linear
and goes from 0 to 2.15 mJy beam$^{-1}$.
{\bf Left.} Zoomed view of GMRT source\#5 coincident with a bright optical/near infrared 
early type star according to CAHA $Ks$-band observations.
This object also has a {\it Chandra} X-ray counterpart indicated by the small
red circle.
{\bf Right.} The same area as in the right panel of Fig. \ref{fig1} containing the bright GMRT source \#3.
This object has an excellent match with one of the 6 cm VLA and {\it Chandra} sources in the subimage. The radio contours
are superimposed onto a WHT $Ks$-band image
showing that there is also a faint near infrared counterpart.
\label{fig2}}
\end{figure*}

VLA-N is well within the $1^{\prime\prime}$ uncertainty of {\it Chandra} \#234 X-ray source and it has also a faint
$Ks=16.8 \pm 0.1$ counterpart in our WHT image (see the right panels of Figs. \ref{fig1} and \ref{fig2}).
Combining VLA+GMRT data, the radio spectrum of this northern VLA source can be described by
$S_{\nu} = (1.35\pm 0.06)~{\rm mJy} [\nu/{\rm GHz}]^{-0.47\pm 0.03}$ and is, therefore,
a clearly non-thermal emitter.
The fit to the {\it Chandra} \#234 spectrum gives a hard photon
index of $1.1\pm 0.4$, unabsorbed flux of $\sim 2\times10^{-13}$~erg~cm$^{-2}$~s$^{-1}$ in the range 2--10~keV, and
hydrogen column density $\sim 2\times10^{22}$~cm$^{-2}$. We note that the data are poor and do not
allow for detailed fitting. Assuming the accepted distance to the Cygnus OB2 association of 1.7 kpc, the X-ray luminosity 
in the range 2--10~keV would be $\sim 10^{32}$~erg~s$^{-1}$. 
Similarly, the 0.1-100 GHz radio luminosity is estimated as $L_{rad} = 1.0 \times 10^{29}$ erg s$^{-1}$.
These values have to be considered as lower limits if the source is beyond the Cygnus OB2 distance.

VLA-S is also of non-thermal nature, with its radio spectrum given by $S_{\nu} = (2.6\pm 0.6)~{\rm mJy} [\nu/{\rm
GHz}]^{-0.5\pm 0.2}$ from VLA data only. 
Undetected by the GMRT above 0.27 mJy ($3\sigma$), its spectral index must be significantly inverted, i.e., with $\alpha \geq
+1.0$ between 49 and 20 cm. This fact is suggestive of synchrotron emission from an optically thick core at long wavelengths. Moreover, this
object also appears superposed onto the {\it Chandra} \#181 error circle 
(see again the right panels of Figs. \ref{fig1} and \ref{fig2}). Since the
number of X-ray counts is too low to perform a proper spectral analysis, we assume a photon index  of 1.5, and hydrogen column density of $\sim
10^{22}$~cm$^{-2}$, and the same distance as above, deriving an X-ray luminosity in the range 2--10~keV of about $10^{31}$~erg~s$^{-1}$. For
the optically thin radio luminosity, we obtain $L_{rad} = 1.6 \times 10^{29}$ erg s$^{-1}$. Inside the {\it Chandra} \#181 circle, there is a
$Ks=13.9\pm 0.1$ near infrared source which is likely of stellar nature based on WHT spectroscopy showing no lines with cosmological redshift.
We propose it to be the counterpart of the X-ray source, provided that the marginal
{\it Chandra} \#181 detection is real. However, the complex morphology of the VLA-S
radio source (see right panel of Fig. \ref{fig1}) suggests that it consists of, at least, two different components barely resolved in the D
configuration of the array. One of them could be the {\it Chandra} X-ray/IR source
and the other the compact radio source detected in our
VLA exploratory observation in A configuration at 3.5 cm. The other one is possibly related to the compact core just commented above, but
further confirmation is yet necessary.  The galactic or extragalactic nature of the mentioned radio core is impossible to establish from
infrared spectroscopy due its extreme faintness in our WHT image ($Ks=19.2 \pm 0.3$).

Another remarkable coincidence is that of the GMRT source \#5, whose position agrees very well with  a $V=11.95$ and $Ks=9.07\pm 0.02$
optical/infrared star belonging to the Cygnus OB2 association. This object is catalogued as star MT91~213 with 
a B0Vp spectral type \citep{mt91,butt2003}. 
There is also the {\it Chandra} \#129 source at the position of such a bright object (see
error circle in the left panel of Fig. \ref{fig2}). With the same assumptions as above, this star has an X-ray luminosity of about
$3\times 10^{31}$ erg s$^{-1}$ in the same energy range. 
From its non VLA detection at 6 cm and excluding variability, 
we obtain a spectral index constrain of $\alpha \leq -0.76$ that points again to a non-thermal
emission mechanism with a radio luminosity of $L_{rad}(0.1-100~{\rm GHz}) = 1.7 \times 10^{28}$ erg s$^{-1}$.
Colliding winds in an early type binary or 
gyrosynchrotron radiation from mildly relativistic electrons trapped in the strong stellar
magnetic field could be responsible for the observed non-thermal emission in this peculiar spectrum star.

We recall that all the luminosities provided here are just lower limits if the sources are further than Cygnus OB2.
In fact, only the GMRT source \#5 is a known member of the association but the others could be more distant.
For instance, at 10~kpc, these objects would present luminosities almost two orders of magnitude larger, 
likely requiring the presence of a compact object, and thus
pointing to an X-ray binary nature (for a model of microquasars -X-ray binaries with jets- powering extended TeV hadronic emission, see
\cite{br2005}).

At this point, we should not forget the existence of the extended source present in the 20 cm VLA map around the TeV CoG. This
feature covers a solid angle of about 27 arc-minute$^2$ and its radius would correspond to $\sim2$ pc at the Cygnus OB2 distance. An
integrated flux density of $116 \pm 3$ mJy is obtained at 20 cm. For synchrotron emission with $\alpha=-1$ at the Cygnus OB2 distance, its
total radio luminosity between 0.1-100 GHz amounts to $L_{rad} \simeq 3.9 \times 10^{30}$ erg s$^{-1}$. From equipartition consideration,
one can also derive a total energy content of $\sim 6.4 \times 10^{45}$ erg and a magnetic field of $1.6 \times 10^{-5}$ G. The
energetics would point to an efficient injector of non thermal particles, coherently with the fact that the region emits at TeV. The
magnetic field value is consistent with a slight enhancement of the interstellar medium density (ISM) value. All this could be explained by a shock
driven by a weakly radiative jet in the ISM produced by some of the point-like sources, perhaps possible microquasars, within the TeV region. In any
case, the radio flux densities reported here provide 
an useful upper limit to the low energy counterpart of the TeV source that (see also \cite{mu2006} for a
diffuse X-ray component inside the TeV source region presenting fluxes of $\sim 10^{-13}$~erg~cm$^{-2}$~s$^{-1}$) allow, and call for,
further constraining the models put forward so far.

To summarize, our results reveal that at least three non-thermal radio sources, one of them being a significantly hard X-ray source,  
and a non-thermal diffuse radio component
are detected in the line of sight towards \tev.  These facts completely change the situation about this unidentified extended TeV source.
Where no counterpart was seen before, now 
several radio sources have emerged and future observations should decide whether or
not one of them is related to the TeV emission. All radio/X-ray sources have significantly negative spectral indices suggestive of optically
thin synchrotron emission. Two of these sources are also of likely stellar origin and one is for sure a bright early type star in Cygnus OB2.
Excluding this last case, the distance to the rest is not yet clearly established. Thus,
their X-ray luminosities may be comparable to those of X-ray binaries if placed beyond Cygnus OB2.
Is one or more of these objects similar to the gamma-ray binaries LS~5039 \citep{par2000, aha2005b} and LS~I+61303 \citep{alb2006}? 
With a farther location in the Galaxy, our sources would have similar 
radio and X-ray luminosities and, if powering the TeV extended
radiation via, e.g., shock with the ISM, the energetic reservoir would be similar as well (see for instance \cite{p2006}). 
An in depth exploration of all these scenarios is beyond the
scope of this letter and will be addressed in a future work.

\acknowledgments
The authors acknowledge support by DGI of the Ministerio de
Educaci\'on y Ciencia (Spain) under grants AYA2004-07171-C02-01 and
AYA2004-07171-C02-02 and FEDER funds. JM is also supported by Plan Andaluz de Investigaci\'on
of Junta de Andaluc\'{\i}a as research group FQM322. GMRT is run by the National Centre for Radio Astrophysics
of the Tata Institute of Fundamental Research.
The National Radio Astronomy Observatory is a facility of the National Science Foundation
operated under cooperative agreement by Associated Universities, Inc.
This paper is also based on
observations collected at the Centro Astron\'omico Hispano Alem\'an
(CAHA) at Calar Alto, operated jointly by the Max-Planck Institut
f\"ur Astronomie and the Instituto de Astrof\'{\i}sica de
Andaluc\'{\i}a (CSIC).
The William Herschel Telescope is operated on the island of La Palma 
by the Isaac Newton Group in the Spanish Observatorio del Roque de los Muchachos of the 
Instituto de Astrof\'{\i}sica de Canarias.



{\it Facilities:} \facility{GMRT ()}, \facility{VLA ()}, \facility{ING:Herschel ()}, \facility{CAO:3.5m ()}

{}

\clearpage



\clearpage


\begin{deluxetable}{ccccc}
\tablecaption{GMRT 610 MHz radio sources above $5\sigma$ peak level.\label{sources}}
\tablehead{
\colhead{Id.}  &  \colhead{$\alpha_{\rm J2000.0}$}  &  
\colhead{$\delta_{\rm J2000.0}$}  &  \colhead{Peak flux} & \colhead{Integrated flux} \\
   &    &     & \colhead{density ($\mu$Jy)} & \colhead{(density $\mu$Jy)} 
}
\startdata
1                  & $20^h 31^m$42\rl 90  & $+41^{\circ} 29^{\prime}$42\pri 7 & $478 \pm 83$ &  $478 \pm 144$   \\
2                  & $20^h 31^m$50\rl 07  & $+41^{\circ} 32^{\prime}$08\pri 4 & $454 \pm 85$ &  $501 \pm 155$   \\
3\tablenotemark{a} & $20^h 31^m$51\rl 78  & $+41^{\circ} 31^{\prime}$18\pri 3 &$1148 \pm 82$ & $1749 \pm 189$   \\
4                  & $20^h 32^m$10\rl 98  & $+41^{\circ} 32^{\prime}$39\pri 6 & $521 \pm 85$ &  $504 \pm 142$   \\
5\tablenotemark{b} & $20^h 32^m$13\rl 09  & $+41^{\circ} 27^{\prime}$24\pri 5 & $525 \pm 82$ &  $718 \pm 178$   \\
6                  & $20^h 32^m$16\rl 29  & $+41^{\circ} 30^{\prime}$55\pri 6 & $566 \pm 82$ &  $772 \pm 176$   
\enddata
\tablenotetext{a}{Coincident with VLA-N, {\it Chandra} \#234 and near IR counterpart.}
\tablenotetext{b}{Coincident with {\it Chandra} \#129 and stellar optical/near IR counterpart,}


\end{deluxetable}


\end{document}